\documentclass[letterpaper]{article} 
\usepackage[]{aaai2027} 
\usepackage[hyphens]{url} 
\usepackage{graphicx} 
\usepackage{natbib} 
\usepackage{caption} 
\usepackage{booktabs}
\usepackage{xcolor}
\usepackage{amsmath,amssymb,amsthm}
\graphicspath{{figures/}}
\newcommand{\method}{\textsc{SHADE}}

\title{SHADE: Shrinkage via Halted Alpha-stable Denoising for Heavy-Tailed Probabilistic Time Series Forecasting}

\author{
Jiahao Weng\textsuperscript{\rm 1},
Jiyu Liu\textsuperscript{\rm 1,\rm 2}
}

\affiliations{
\textsuperscript{\rm 1}Renmin University of China\\
\textsuperscript{\rm 2}Kuaishou Technology
}

\begin{document}
\maketitle

\begin{abstract}
Time-series observations are particularly susceptible to noise, and many real-world series have low signal-to-noise ratios (SNRs) together with heavy-tailed and time-varying disturbances. When Gaussian diffusion models are used for forecasting, reversing all the way to the zero-noise endpoint may yield an overly narrow and overconfident predictive distribution. We propose SHADE, an alpha-stable L\'evy diffusion forecaster that reads out at a nonzero depth so that the residual stable kernel acts as distribution-level shrinkage. By decomposing halting risk into smoothing bias and estimation error, we develop a scale-calibrated halt for estimating the optimal shrinkage $\tau^*$ and a context-amortized halt that requires no explicit noise labels. Gains concentrate in the extreme heavy-tailed regime $\alpha\leq1.3$: Across synthetic noise settings, median CRPS improves by 21.7\%, while the NLL falls significantly; Cryptocurrency and foreign-exchange data provide further evidence of improved tail calibration in real-data implement. On near-light-tailed or high-SNR data, the model reverts to Gaussian diffusion or full denoising. SHADE maintains overall forecast error comparable to strong baselines while providing more honest tail uncertainty for heavy-tailed series.
\end{abstract}

\section{Introduction}

Unlike text or images, which are often modeled as observations near a high-dimensional manifold, a time series is a sequence of repeated measurements of an evolving latent state. A clean, error-free trajectory is rarely observed. For $x_t=s_t+\varepsilon_t$, we usually use the physically interpretable scale ratio such as $\mathrm{SNR}=\mathbb{V}\mathrm{ar}(s_t)/\mathbb{V}\mathrm{ar}(\varepsilon_t)$ \cite{moran2018eegsnr,herreraparedes2018communities,lim2024semgdataset}; SNR is a vital influence factor in time series forecast. Noise is also domain specific: motion, electrode displacement, power-line interference, and sensor error contaminate biological or industrial measurements \cite{kalita2024aneeg,liu2024manufacturing}, while weak financial predictability is obscured by market microstructure and event shocks \cite{cont2001stylized,aitsahalia2005microstructure}. At coarse scales, many perturbations may aggregate toward a Gaussian approximation. But at fine resolution, however, fewer shocks are averaged within each observation and individual jumps remain visible; the effective disturbance can therefore be non-Gaussian, heavy-tailed, clustered, and time varying.

Machine learning methods, like state-space filters, wavelets, and empirical-mode decompositions separate noise through dynamics, time-frequency locality, or intrinsic oscillations; learned denoisers and LLM-based representations add flexible nonlinear mappings \cite{langarica2023blind,zhang2024llmts,zhang2024ssl}. Even simple linear predictors remain surprisingly competitive \cite{zeng2023linear}. Such pipelines commonly choosing a cleaned trajectory by MSE, correlation are naturally designed for point estimation. Nevertheless, a point estimate collapses multiple plausible futures and can erase randomness that is itself decision relevant. The same conditional mean may accompany radically different tail probabilities; low MSE guarantees neither calibrated interval coverage nor reasonable density at an extreme observation \cite{gneiting2011point}. In risk limits, predictive maintenance, disaster warning, and climate adaptation, underestimating a tail event can cost more than a modest median error. Our target is therefore a calibrated yet sharp conditional distribution of future observations, not a single denoised trajectory.

\begin{figure*}[htbp]
\centering
\includegraphics[width=0.97\textwidth]{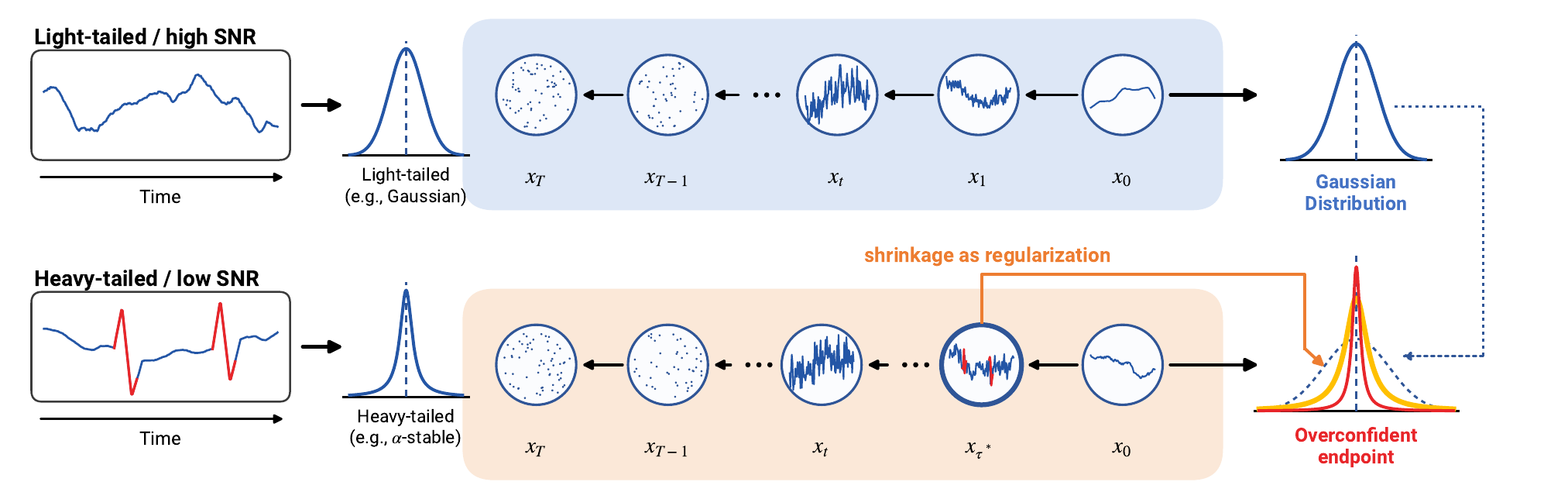}
\caption{Why alpha-stable diffusion and halted shrinkage. A Gaussian/light-tail denoiser poorly matches jump-like observations, while LIM supplies an alpha-stable reverse process whose jumps and scale aggregation match heavy-tailed behavior \cite{yoon2023levy}. Full reversal of LIM which is designed for CV, however, removes all stable smoothing is unreliable for time series probabilistic estimation. \method{} therefore reads at $\tau>0$: the residual multiplier $\exp[-\gamma(\tau)^\alpha|\omega|^\alpha]$ suppresses unstable frequencies but retains more genuine structure than excessive smoothing. This is distributional shrinkage---a bias--error trade-off related to James--Stein estimation and regularized early stopping \cite{james1961estimation,yao2007early,raskutti2014early}---rather than a computational shortcut.}
\label{fig:heavy-tail}
\end{figure*}

Score-based diffusion offers a flexible route to this distributional objective. Its continuous-time view evolves densities through a Fokker--Planck equation and transports a noise law back toward data with a time-dependent score. TimeGrad, CSDI, Diffusion-TS, Dynamical Diffusion, and Non-stationary Diffusion develop autoregressive conditioning, time--variable attention, structural generation, dynamic consistency, and time-varying uncertainty \cite{rasul2021timegrad,tashiro2021csdi,yuan2024diffusionts,guo2025dydiff,ye2025nsdiff}; Noise2Score further connects a score learned from noisy marginals to posterior statistics \cite{kim2021noise2score}. While most time-series variants combine Gaussian forward corruption with a fixed-step-and-zero-noise readout, Gaussian diffusion is not intrinsically mis-calibrated. But under heavy-tailed observations and finite training data, the high-frequency score becomes hard to estimate. Over-reversal denoise can then amplify local score error, yielding a distribution that remains accurate near its center but is too narrow in its tails and assigns vanishing density---hence extreme NLL---to rare observations.

We propose \method{} (Shrinkage via Halted Alpha-stable Denoising), which selects a nonzero readout with a strictly proper distribution score \cite{gneiting2007proper}. In theoretical field, we conduct that the score's excess risk becomes the squared distance between predictive and true CDFs, enabling an exact spectral bias--error decomposition and a locally explicit SNR--depth law. In Algorithm, an auditable route (M1, M2) maps tail-index and SNR scale to halting depth, while a context-amortized route learns the depth distribution without unobservable noise labels. We test both routes on a complete tail-index--SNR grid and five real data domains, and organize ablation experiment about noise-family misspecification, equal-width counterfactuals. Our contributions are:
\begin{enumerate}
\item We formulate conditional alpha-stable forecasting with nonzero readout as distributional shrinkage for low-SNR heavy-tailed series.
\item We derive a spectral bias--error decomposition as fundamental, with complementary scale-based and context-amortized halting.
\item An improved tail uncertainty estimating method on heavy-tail data regime is presented, without reducing overall estimation accuracy significantly.
\end{enumerate}

\section{Related Work}

\subsection{Score-Based Probabilistic Time-Series Forecasting}

Diffusion forecasters progressively enrich conditioning: TimeGrad uses recurrent autoregressive sampling, CSDI attends over time and variables, and Diffusion-TS, Dynamical Diffusion, and Non-stationary Diffusion introduce structural decomposition, explicit dynamics, and adaptive location--scale schedules \cite{rasul2021timegrad,tashiro2021csdi,yuan2024diffusionts,guo2025dydiff,ye2025nsdiff}. They largely retain Gaussian corruption and zero-noise readout. Related noisy-learning methods recover posterior statistics without clean targets, select denoising in representation space, or improve financial downstream signals \cite{kim2021noise2score,zhou2024decl,wang2024financial}; these still emphasize a reconstructed trajectory. We instead select a full conditional distribution with a strictly proper score, making halting part of estimation rather than preprocessing.

\subsection{Non-Gaussian Diffusion Models}

Non-Gaussian diffusion has used Gamma mixtures, generalized-normal corruption, and Student-$t$ transitions to improve mode or rare-event coverage \cite{nachmani2021nongaussian,deasy2021heavytail,pandey2025heavytail}. Stable L\'evy motion supplies a continuous-time route: fractional Langevin sampling crosses low-density barriers, LIM derives fractional score matching and exact reversal, and DLPM shows that discrete stable diffusion has a distinct backward process \cite{simsekli2017fractional,yoon2023levy,shariatian2025dlpm}. Markov jumps, Poisson thinning, and Brownian--jump mixtures likewise support categorical, sparse-count, and finite-activity generation \cite{sun2022discrete,chen2023jump,baule2025jumpdiffusion}. These works establish valid non-Gaussian kernels; we study the statistical meaning and data-dependent selection of a residual stable kernel.

\section{Mathematical Background}

\subsection{Alpha-Stable L\'evy Processes and Reverse Dynamics}

L\'evy processes combine stationary independent increments with c\`adl\`ag paths, so they can describe both the accumulation of small perturbations and isolated jumps. We use the symmetric alpha-stable subclass because it is closed under addition and rescaling and is the generalized-central-limit attractor of heavy-tailed sums. Consequently, repeated aggregation changes its scale but not its distributional family. For a unit-scale driver,
\[
\mathbb E\exp(i\xi L_t^\alpha)=\exp\!\left(-t|\xi|^\alpha\right).
\]
A scale $\gamma_L$ multiplies the exponent by $\gamma_L^\alpha$. The tail index $\alpha$ determines how frequently large increments occur: $\alpha=2$ is the Gaussian boundary, while smaller values produce progressively heavier tails. Symmetry separates this tail parameter from scale and leaves location and directional structure to the conditional network. We restrict $\alpha\in(1,2]$ so the integrated distribution score below has the required first absolute moment.

An ordinary spatial score is local, whereas jumps generate nonlocal density evolution. Following LIM, the corresponding fractional score is therefore defined through the Riesz operator
\[
\mathcal G_\alpha=(-\Delta)^{(\alpha-2)/2}\nabla,
\qquad
S_p^{(\alpha)}(x)=\frac{\mathcal G_\alpha p(x)}{p(x)}.
\]
It reduces continuously to $\nabla\log p$ at $\alpha=2$. For the variance-preserving process specified in Section~\ref{sec:method}, LIM's time-reversal theorem gives the following reverse dynamics, integrated from $T$ toward $0$:
\[
\begin{aligned}
dX_t={}&\left[-\frac{\beta(t)}{\alpha}X_t
-\alpha\beta(t)S_t^{(\alpha)}(X_t)\right]dt+\beta(t)^{1/\alpha}d\bar L_t^\alpha.
\end{aligned}
\]
Here the first drift term reverses the linear attenuation, the fractional-score correction steers mass toward the data law, and $\bar L_t^\alpha$ is the reverse stable driver. This expression is the jump-process analogue of the Gaussian reverse SDE. The closed-form transition, fractional Fokker--Planck equation, and specialization of the reversal theorem are given in Appendix~A.

\paragraph{Rationale for model matching.} Artificial diffusion noise is a training device, whereas observation noise belongs to the unknown data-generating process; equality between them is neither observable nor required. Matching their broad behavior---tail index, jump frequency, and scale aggregation---is nevertheless useful: one stable jump can represent an extreme shock without explaining it as an unlikely accumulation of Gaussian steps. It also gives the training target, reverse trajectory, and residual kernel a common scale calculus. Varying depth then changes the amount of smoothing without switching tail family: $\alpha$ controls kernel geometry and $\gamma_t$ its strength. Stable closure thus provides a continuous regularization path even for approximately stable, heteroscedastic, or Student-$t$ disturbances.

\subsection{Distributional Risk and an Explicit Halting Law}

We require a risk that compares complete predictive distributions without assuming a finite variance. For a predictive CDF $F$, define $\mathcal S(F,y)=\int_{\mathbb R}(F(z)-\mathbf1\{y\leq z\})^2dz$. If $P$ is the true conditional law and $Q$ is a forecast, strict propriety gives
\[
\begin{aligned}
&\mathbb E_{Y\sim P}\mathcal S(F_Q,Y)
-\mathbb E_{Y\sim P}\mathcal S(F_P,Y)\\
&\qquad=\int_{\mathbb R}
\left(F_Q(z)-F_P(z)\right)^2dz.
\end{aligned}
\]
The excess risk is exactly the squared $L_2$ distance between CDFs. Halting is therefore a distribution-estimation problem that simultaneously sees the center, quantiles, and cumulative tails, rather than a rule for choosing one point forecast.

Write $s=\mathrm{SNR}$ and represent the random fully denoised estimate as $\widehat F(\cdot;s)=F_P+\delta(\cdot;s)$. The error $\delta$ collects finite-sample variation, network approximation, score error, and optimization randomness; it is not the observation noise itself. Reading at $\tau>0$ convolves this estimate with a stable kernel. Its residual width is $r=\gamma(\tau)$, and the corresponding spectral semigroup time is $\lambda=r^\alpha$:
\[
\widehat K_r(\omega)=\exp\!\left(-\lambda|\omega|^\alpha\right).
\]
The three quantities have distinct interpretations: $\tau$ is reverse-process depth, $r$ is smoothing width, and $\lambda$ is the natural time parameter of the stable multiplier. Below, let $w=|\omega|^\alpha$. We write $A(\omega)=|\widehat F_P(\omega)|^2$ for the true CDF's structural spectrum and $E(\omega,s)=\mathbb E|\widehat\delta(\omega;s)|^2$ for the endpoint estimation-error spectrum.

Assuming $\mathbb E\delta=0$ removes the cross term. Parseval's identity then yields the exact bias--error decomposition
\[
\begin{aligned}
R(\lambda;s)&=B(\lambda)+V(\lambda;s),\\
B(\lambda)&=\int
\left|1-e^{-\lambda w}\right|^2
A(\omega)\,d\omega,\\
V(\lambda;s)&=\int
e^{-2\lambda w}E(\omega,s)\,d\omega.
\end{aligned}
\]
$B$ measures genuine CDF structure removed by the residual kernel. It starts from zero and grows with smoothing. In contrast, $V$ is the estimation error that survives the same kernel and decreases most rapidly at high frequency. A shallow halt can therefore remove unstable local variation before substantially damaging the broad distributional shape. Their marginal balance may have an interior optimum; if endpoint error is already weak, the solution returns to $\tau=0$. Exact derivatives, boundary conditions, and the nonzero-mean cross term are deferred to Appendix~B.

\subsubsection{Inspire an explicit estimable halting law.}

The exact balance still depends on two unknown spectra. To obtain an estimable law, we approximate only a neighborhood of an interior optimum rather than imposing a power law on the complete trajectory. If the local log-slopes are stable, first-order expansion in $(\log\lambda,\log s)$ gives the testable regular-variation form
\[
B(\lambda)\simeq C_B\lambda^p,
\qquad
V(\lambda;s)\simeq C_Vs^{-q}\lambda^{-r}.
\]
$C_B$ and $C_V$ are local reference energies, while $p,r,q>0$ respectively describe bias growth, error attenuation, and SNR sensitivity. These quantities can be checked from neighboring depths and held-out SNR regimes. The approximation is local and must not be extrapolated to $\lambda=0$. Minimizing this local risk gives
\[
\lambda^*(s)=\left(\frac{rC_V}{pC_B}\right)^{1/(p+r)}s^{-q/(p+r)}.
\]
If the schedule is locally self-similar around the optimal interval, $\gamma(\tau)\simeq G\tau^H$, then $\lambda=\gamma(\tau)^\alpha\simeq G^\alpha\tau^{\alpha H}$. Inverting the schedule gives
\[
\begin{aligned}
\tau^*&\simeq c(\alpha)\,
\mathrm{SNR}^{b(\alpha)},\\
b(\alpha)&=-\frac{q}{\alpha H(p+r)}<0,\\
c(\alpha)&=G^{-1/H}
\left(\frac{rC_V}{pC_B}\right)^{\!1/[\alpha H(p+r)]}.
\end{aligned}
\]
The negative exponent formalizes the intended behavior: lower SNR calls for an earlier halt, while improved score estimation lowers $C_V$ and moves $\tau^*$ toward zero. The exponent is not universal because $\alpha$ changes both the stable low-pass geometry and the local spectral slopes. If the slopes are unstable or the solution lies outside the candidate interval, we use numerical calibration or context amortization instead. Appendix~B supplies the skipped derivatives, existence and boundary results, approximation diagnostics, and the biased-error extension.

\section{Method}\label{sec:method}

\subsection{Problem Definition}

Given a $D$-variate series $x_{1:T}$, at time $t$ the model uses a history window of length $L$, $c_t=x_{t-L+1:t}$, to predict the conditional distribution of the next $H$ observations, $y_0=x_{t+1:t+H}$. Training, halting calibration, and testing are split in chronological order. Because variance may not exist for $\alpha<2$, every sample is normalized only with the median and median absolute deviation of its history window; predictive samples are then transformed back to the original scale.

\method{} contains a conditional alpha-stable denoiser and four functional halting components. In denoiser, the history window is treated as conditional observation and the future window as prediction target; forward diffusion perturbs only the target, while observations remain fixed at every denoising step. But unlike CSDI, the condition includes not only history, mask, and temporal--variable position encodings but also a FiLM embedding $\psi(\alpha)$ of the tail index. The network therefore learns
\[
s_\theta\!\left(y_k,k\mid c_t,m,\psi(\alpha)\right),
\]
where $y_k$ is the noisy target at step $k$, $m$ distinguishes conditional and predictive locations, and $\psi(\alpha)$ specifies the tail thickness to which the denoiser should adapt. Synthetic data supply $\alpha$ from the generator; in single-model real-data experiments, $\alpha$ is estimated only on the training segment and is never adjusted using test scores.

\subsection{Conditional Denoising}

\subsubsection{Forward Perturbation}

The continuous forward process perturbs only the future target:
\[
\begin{aligned}
dY_t={}&-\frac{\beta(t)}{\alpha}Y_t\,dt
+\beta(t)^{1/\alpha}dL_t^\alpha,\\
Y_t={}&a_tY_0+\gamma_t\varepsilon,
\qquad \varepsilon\sim S\alpha S(1).
\end{aligned}
\]
where $a_t=\exp[-\alpha^{-1}\int_0^t\beta(u)du]$ and $\gamma_t=(1-a_t^\alpha)^{1/\alpha}$. Its analytic conditional fractional score is
\[
S_t^{(\alpha)}(Y_t\mid Y_0)=-\frac{\varepsilon}{\alpha\gamma_t^{\alpha-1}}.
\]
On $k=0,\ldots,K$, let $\vartheta_k=((k/K+s)/(1+s))\pi/2$. The cosine schedule is written compactly as $a_k=\cos(\vartheta_k)/\cos(\vartheta_0)$; we then set $\gamma_k=(1-a_k^\alpha)^{1/\alpha}$ and sample $y_k=a_ky_0+\gamma_k\varepsilon$. We take $K=50$ and $s=0.008$; training samples one $k$ uniformly without simulating the full chain. History $c_t$ remains fixed, so the model learns a conditional distribution rather than reconstruction.

\subsubsection{Denoising Network with FiLM-$\alpha$ Conditioning}

The backbone follows CSDI's residual design and alternates attention over time and variables in every block to capture temporal dependence and shared shocks. The noisy target, history, mask, and depth embedding are first mapped to hidden features. Instead of concatenating a low-dimensional tail index to the input plane, FiLM generates a channel-wise scale and shift for residual block $\ell$:
\[
\begin{aligned}
&e_\alpha=\psi(\alpha)
=\mathrm{MLP}_\alpha(\phi(\alpha)),\\
&(r_\ell,b_\ell)=W_\ell e_\alpha,\\
&h_\ell\leftarrow r_\ell\odot
\mathrm{Norm}(h_\ell)+b_\ell.
\end{aligned}
\]
$\phi(\alpha)$ normalizes the tail index to a fixed interval, while $r_\ell$ and $b_\ell$ are learned. FiLM makes identical histories induce different internal representations at different $\alpha$, so the tail index is part of the conditional law rather than merely an external forward-noise setting.

\subsubsection{Training Loss and Reverse Denoising}

\method{} predicts injected alpha-stable noise and uses the analytic fractional score of the stable kernel as LIM \cite{yoon2023levy}. For a sampled window, depth $k$, and stable noise $\varepsilon$, it minimizes
\[
\begin{aligned}
\mathcal L_{\mathrm{SHADE}}(\theta)
={}&\mathbb E\!\Big[
\big\|
\gamma_k^{\alpha-1}
s_\theta\!\left(y_k,k\mid c_t,m,\psi(\alpha)\right)
+\frac{\varepsilon}{\alpha}
\big\|_1\Big].
\end{aligned}
\]
The target follows from the conditional fractional score; $\gamma_k^{\alpha-1}$ puts different depths on a comparable scale. We use $L_1$ rather than squared loss because $\alpha\in(1,2]$ guarantees only a first absolute moment, making $L_2$ unreliable under extreme stable samples.

At inference, the predictive locations are initialized with symmetric stable noise while history remains fixed. The network repeatedly predicts and removes current noise, saving a predictive state at every candidate depth. A fully denoised model runs to $k=0$; \method{} differs only in allowing the components below to select $\tau\geq0$. The continuous reverse L\'evy process and its conditions have already been given in the theory section.

\begin{figure*}[htbp]
\centering
\includegraphics[width=0.97\textwidth]{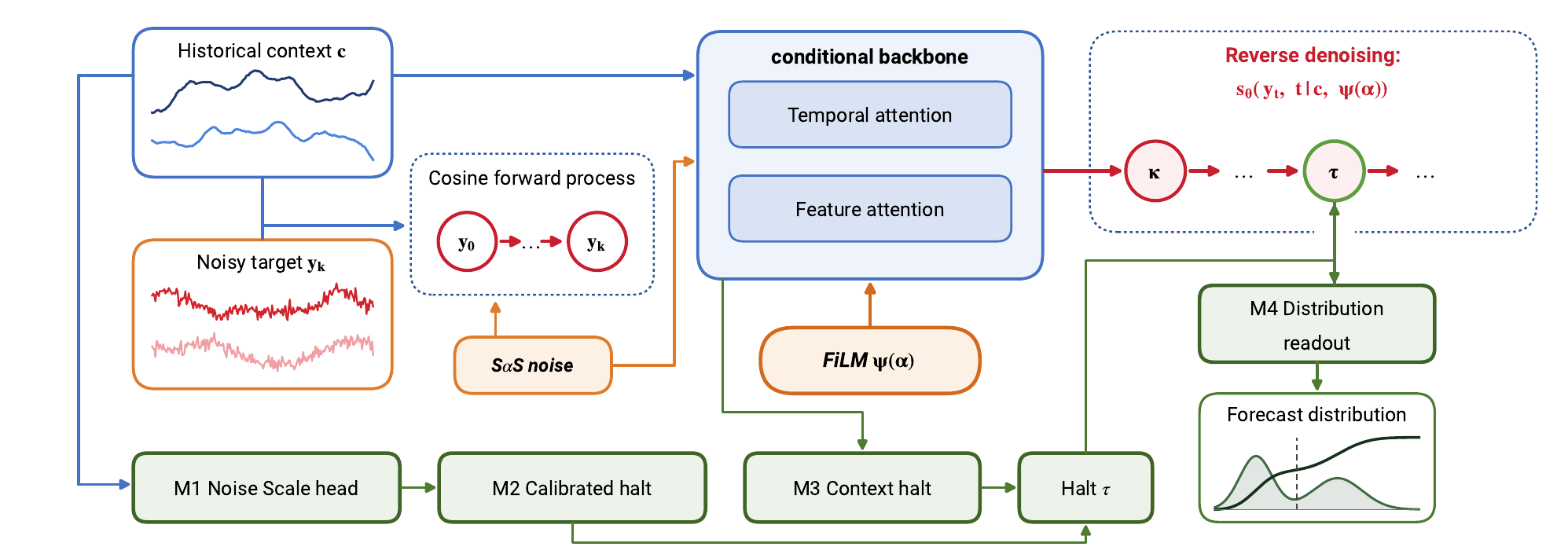}
\caption{Architecture and four functions of a single \method{}. The time--variable backbone is FiLM-modulated by $\psi(\alpha)$. The robust scale estimator summarizes level, adjacent-difference, and seasonal-difference MADs; the spectral rule maps that scale to the calibrated residual width $C_\alpha\widehat\gamma_z^{q_\alpha}$; alternatively, the context policy maps pooled history and $\psi(\alpha)$ to a distribution over depths without noise labels. The shared reader saves candidate states along identical stable reverse trajectories and converts the selected state into an empirical CDF and intervals. The denoiser is trained first, then frozen while the two halting routes are fitted on chronological calibration data; neither route sees test futures.}
\label{fig:shade-architecture}
\end{figure*}

\subsection{Halting Components}

The four functions in Fig.~\ref{fig:shade-architecture} are constructed and trained as follows.

\paragraph{Robust local noise-scale estimator (M1).}
This component supplies the interpretable route with a window-level disturbance scale. Its input $z(c_t)$ contains the MAD of the observed level, adjacent differences, and, when available, seasonal differences; these summarize total scale, rapid variation, and residual variation without relying on a finite variance. Together with the tail-index embedding, a small positive regression head predicts
\[
\begin{aligned}
\widehat\gamma_z(c_t)
=\exp\!\Big[
f_{\mathrm{scale}}\!\big(
z(c_t),\psi(\alpha)
\big)\Big].
\end{aligned}
\]
Known synthetic noise scales provide direct supervision. On real data, only training-segment high-frequency differences provide a weak target, so this estimator remains an auditable route rather than an assumption that every difference is pure noise.

\paragraph{Spectrally calibrated halting rule (M2).}
This component converts the estimated scale into a residual diffusion width instead of equating it mechanically with observation noise. A chronological calibration set fits $C_\alpha$ and $q_\alpha$, after which the rule selects the scheduled depth whose residual scale $g_k$ is nearest to the calibrated target:
\[
\begin{aligned}
r^*(c_t)
&=C_\alpha\widehat\gamma_z(c_t)^{q_\alpha},\\
\widehat\tau_{\mathrm{spec}}(c_t)
&=\arg\min_k
\left|g_k-r^*(c_t)\right|.
\end{aligned}
\]
The denoiser is frozen during this calibration, and all constants are fixed before testing; consequently, every window-level depth can be traced back to an observed robust scale.

\paragraph{Context-amortized halting policy (M3).}
Because signal changes and noise changes are difficult to separate from differences alone, this alternative bypasses an explicit noise label. It pools the frozen CSDI context representation, $u_t=\mathrm{Pool}(H_\theta(c_t,m))$, and maps it with $\psi(\alpha)$ to a probability over candidate depths:
\[
\begin{aligned}
\pi_\eta(k\mid c_t,\alpha)
=\mathrm{softmax}\!\Big[
f_{\mathrm{halt}}\!\big(
u_t,\psi(\alpha)
\big)\Big]_k.
\end{aligned}
\]
Depth-wise predictive CDFs are cached on calibration windows, and only the policy parameters $\eta$ are trained with their mixture's integrated distribution score. No observation-noise label or test future is used; when shrinkage is unnecessary, the policy can place its mass at $k=0$.

\paragraph{Shared-trajectory distributional reader (M4).}
Multiple reverse trajectories are drawn once and their attenuated centers $v_t^{(s)}$, instead of raw reverse state, are saved at candidate depth. Let $m_c$ and $d_c$ be the history median and MAD. Since $Y_t=a_tY_0+\gamma_t\varepsilon$, and a stable residual can be written as $\varepsilon=\sqrt{A}Z$ with $Z\sim\mathcal N(0,1)$ and  $A\sim$ positive $\alpha/2$-stable distribution \cite{samorodnitsky1994stable}, the reader evaluates
\[
\begin{aligned}
\mu_{t,s}&=m_c+d_c\frac{v_t^{(s)}}{a_t},\quad \rho_t=d_c\frac{\gamma_t}{a_t},\\
\widehat f_t(y)
&=\frac1S\sum_{s=1}^{S}\mathbb E_{A_s}\left[
\varphi\!\left(y;\mu_{t,s},\rho_t^2A_s\right)\right].
\end{aligned}
\]
Thus, here $\varphi(\cdot;\mu,\sigma^2)$ is a Gaussian density. The first line remove VP attenuation and show that $\gamma_t/a_t$, not $\gamma_t$ alone, is the residual scale on data coordinates. The mixture analytically retains the stable kernel; it adds neither a second perturbation nor a post-hoc variance multiplier, and supplies the CDF, quantiles, intervals, and NLL. Both routes reuse identical trajectories, isolating depth from resampling. Training is staged: fit the FiLM denoiser, then freeze it while calibrating the two halting routes.

\section{Experiments}

We report NLL, CRPS, and 90\% coverage; and NLL is primary because it directly penalizes vanishing density at a realized tail event, whereas CRPS averages CDF error over the line. For \method{}, window-level NLL is $-\log\widehat f_{\tau^*}(y)$ under the M4's Gaussian scale-mixture density above, and we report its mean value. Analytic baselines use their own likelihoods; sample-only baselines use a common KDE bandwidth. Both deployable halting routes are fitted on the chronological calibration segment; unless reported separately, $\tau^*$ denotes the depth returned by whichever route---scale estimation plus spectral calibration or context amortization---has lower calibration CRPS. More detailed information and results about experiment are disclosed in appendix D and E.

\subsection{Synthetic Experiments}

\subsubsection{Generator and Parameters}

We first construct a controlled signal-plus-noise process. Let
\[
\begin{aligned}
u_t&=\sum_{j=1}^{p}\phi_j u_{t-j}+\epsilon_t,\\
x_t&=u_t+z_t,
\qquad z_t\sim S\alpha S(\gamma_z).
\end{aligned}
\]
where the autoregressive polynomial is stable, $\epsilon_t$ drives predictable-state innovations, and $z_t$ is an independent symmetric alpha-stable observation disturbance with characteristic function $\exp(-\gamma_z^\alpha|\xi|^\alpha)$. The tail index $\alpha$ controls extreme-jump frequency and $\gamma_z$ controls disturbance strength. Defining $\mathrm{SNR}=\gamma_u/\gamma_z$ with robust latent scale $\gamma_u$ separates changes in latent dynamics from changes in observation-tail thickness and strength.

We set $p=1$, $\phi_1=0.9$, and use Gaussian state innovations. The grid combines $\alpha\in\{1.1,1.2,\ldots,2.0\}$ with $\mathrm{SNR}\in\{0.05,0.1,0.2,0.5,1,2,5,10\}$, producing 80 settings. Each uses training median--MAD normalization, context 32, horizon 8, 50 diffusion steps, and a common predictive-sample budget. The boundary $\alpha=2$ is Gaussian; $\alpha\leq1.3$ is designated heavy-tailed and $1.4$--$1.5$ transitional before observing outcomes.

All 80 cell-wise metrics and boundary optima are reported in the appendix.

\subsubsection{Tail-Index--SNR Grid}

\begin{figure}[htbp]
\centering
\includegraphics[width=\columnwidth]{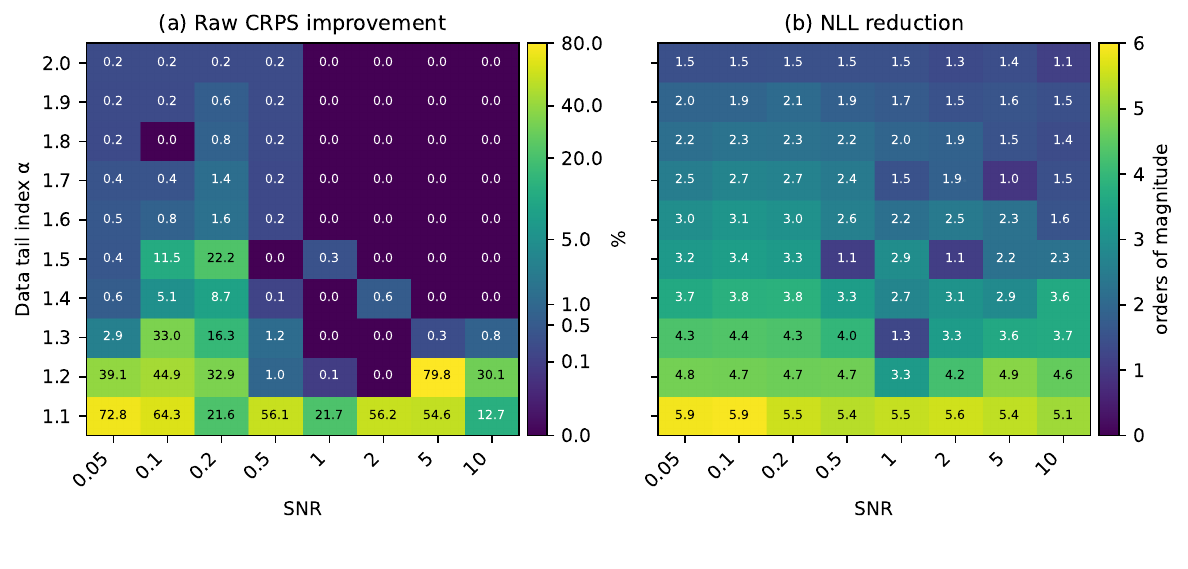}
\caption{Relative CRPS improvement and orders-of-magnitude NLL ratio over the complete 80-cell synthetic grid.}
\label{fig:synthetic-heatmap}
\end{figure}

Figure~\ref{fig:synthetic-heatmap} shows relative CRPS gain and $\log_{10}(\mathrm{NLL}_{0}/\mathrm{NLL}_{\tau^*})+1$; complete cell-wise metrics are reported in the appendix.

The benefit concentrates exactly where hypothesized: for $\alpha\leq1.3$, median CRPS improves 21.7\% and max NLL falls 4.69 orders, whereas CRPS is essentially unchanged beyond the transition region. NLL remains more responsive because a few tail observations receive catastrophically small endpoint density even when their contribution to integrated CDF error is modest.

\subsubsection{Comparison with Conditional Diffusion Baselines}

We compare methods on random pooled synthetic heavy-tail and low-SNR data ($\alpha \leq 1.5$, $\mathrm{SNR} \leq 1$) and 4,500 evaluation windows to see its concrete performance. CSDI stands for Gaussian ($\alpha=2$) diffusion; LIM stands for an alpha-stable process with full endpoint denoising ($\tau = 0$). The two deployable \method{} variants use either robust scale estimation followed by spectral calibration or a context-amortized distribution over depths. Evaluation windows and metric definitions are identical.

\begin{table}[htbp]
\centering\footnotesize
\resizebox{\columnwidth}{!}{%
\begin{tabular}{lccc}
\toprule
Method & CRPS  & NLL  & 90\% coverage \\
\midrule
CSDI (Gaussian, $\alpha=2$) & 9.45 & 1.44 & 0.855 \\
LIM (full denoising, $\tau = 0$) & 10.15 & 111.26 & 0.876 \\
\method{} (M1, M2) & 8.87 & 1.46 & 0.919 \\
\method{} (M3) & \textbf{8.73} & \textbf{1.13} & \textbf{0.910} \\
\bottomrule
\end{tabular}}
\caption{Comparison on pooled heavy-tailed synthetic data.}
\label{tab:synthetic-baselines}
\end{table}

Table~\ref{tab:synthetic-baselines} confirms that stable noise alone is insufficient: full-denoising LIM is overconfident, whereas either nonzero readout restores density and near-nominal coverage. Context amortization is slightly stronger because it adapts within a nominal noise regime without unobserved window-level noise labels.

\subsection{Real-Data Experiments}

Cause the tail indices of real data are different from group to group and time-varying, we first used the McCulloch method to statistically estimate its tail index alpha, and then embedded it into the \method{} model through FiLM.

\subsubsection{Data and comparison methods choosing}

To examine whether the regime identified in simulation persists in natural data, we evaluate five domains spanning markedly different tail behavior and temporal structure. They comprise hourly closing prices for six cryptocurrencies (BTC, ETH, BNB, SOL, XRP, and DOGE), the eight daily foreign-exchange series in GluonTS, 120 hourly GluonTS electricity-consumption series, 120 hourly GluonTS traffic-occupancy series \cite{alexandrov2020gluonts}, and daily NG, CL, HO, RB, and BZ energy futures from Yahoo Finance.

We choose comparison methods to cover distinct probabilistic assumptions and forecasting architectures rather than to claim an exhaustive leaderboard. Gaussian MLE is a low-capacity parametric reference; DLinear-Prob and PatchTST represent linear and patch-based Transformer forecasters \cite{zeng2023linear,nie2023patchtst}; TimeGrad is an established autoregressive diffusion forecaster \cite{rasul2021timegrad}; and CSDI is the closest controlled diffusion baseline \cite{tashiro2021csdi}. All methods receive the same preprocessed pools, context lengths, horizons, and evaluation targets. Dataset-specific windows, the split method, SOTA model and training devices data are detailed in Appendix D.

\begin{table*}[hbtp]
\centering\footnotesize
\setlength{\tabcolsep}{3.2pt}
\resizebox{\textwidth}{!}{
\begin{tabular}{lccc*{6}{cc}}
\toprule
Dataset & \multicolumn{2}{c}{\method{}} & \multicolumn{1}{c}{mean of SOTA} & \multicolumn{2}{c}{Gaussian MLE} & \multicolumn{2}{c}{DLinear-Prob$^{(2023)}$}  & \multicolumn{2}{c}{TimeGrad$^{(2021)}$} & \multicolumn{2}{c}{PatchTST$^{(2023)}$} & \multicolumn{2}{c}{CSDI$^{(2021)}$} \\
\cmidrule(lr){2-3} \cmidrule(lr){4-4} \cmidrule(lr){5-6} \cmidrule(lr){7-8} \cmidrule(lr){9-10} \cmidrule(lr){11-12} \cmidrule(lr){13-14} 
 & NLL & CRPS & CRPS & NLL & CRPS  & NLL & CRPS & NLL & CRPS & NLL & CRPS & NLL & CRPS \\
\midrule
Foreign exchange ($\bar\alpha\approx1.31$) & $\color{red}\mathbf{1.365}$ & 0.7062 & 0.7037 & $\inf$ & $\inf$ & 1.642 & 0.7132 & $2.168^{\dagger\dagger}$ & 0.6969 & $\color{blue}1.590$ & 0.6996 & 2.509 & 0.7052 \\
Crypto ($\bar\alpha\approx1.42$) & $\color{red}\mathbf{1.534}$ & 0.8960 & 0.8815 & 1.866 & $0.8852$ & 1.816 & 0.9045 & $2.232^\dagger$ & 0.8795 & $\color{blue}1.687$ & 0.8609 & 2.736 & 0.8812 \\
Electricity ($\bar\alpha\approx1.69$) & $\color{red}\mathbf{0.334}$ & 0.3042 & 0.3335 & $1.168^\dagger$ & 0.4382 & 0.647 & 0.2821 & $1.689^{\dagger\dagger}$ & 0.3592 & 0.453 & 0.2822 & $\color{blue}0.386$ & 0.3056 \\
Traffic ($\bar\alpha\approx1.84$) & $\color{blue}{-0.143}$ & 0.2917 & 0.3008 & 0.464 & 0.2862 & 0.622 & 0.3385 & $1.324^{\dagger\dagger}$ & 0.3207 & -0.056 & 0.2692 & $\color{red}\mathbf{-0.141}$ & 0.2894 \\
Energy futures ($\bar\alpha\approx2.00$) & $\color{blue}-0.967$ & 0.0689 & 0.0691 & $-0.765^\dagger$ & 0.0675 & $-0.749^\dagger$ & 0.0685 & $-0.577^\dagger$ & 0.0738 & -0.817 & 0.0665 & $\color{red}\mathbf{-1.027}$ & 0.0693 \\
\bottomrule
\end{tabular}}
\caption{Unified probabilistic comparison on five real datasets. NLL is the primary metric and appears to the left of CRPS. Bold red and blue denote the best and second-best NLL, respectively. $^\dagger$ and $^{\dagger\dagger}$ notions present SE$\geq$ 5\% NLL and SE $\geq$ 10\% NLL respectively, and the other satisfy SE $\leq$ 5\% NLL.}
\label{tab:real-data}
\end{table*}

\subsubsection{Comparison with Recent Probabilistic Forecasters}

Table~\ref{tab:real-data} shows NLL as primary and CRPS as a supporting indicator reflecting overall performance, with the best and second-best NLL shown in bold red and blue. Gaussian MLE is non-finite on foreign exchange and is retained as inf.

The ordering matches our hypothesis: \method{} has the best NLL on the heaviest-tailed return domains, and loses its density advantage as $\bar\alpha$ approaches two, but still keeps competitive on Energy futures which is almost light-tail data. Compared with the average CRPS of SOTA, the \method{} maintains a similar level. The \method{} achieved a tail prediction improvement of over 9.1\% on NLL with a difference of less than 1.65\% on heavy tailed data (FX and crypto). And \method{} performs even better than average on the electricity, traffic, and energy futures datasets. Per-dataset diagnostics are reported in the appendix.

\subsection{Ablation Studies}

\subsubsection{Equal-IQR Variance-Inflation Counterfactual}

The most direct alternative explanation is that any wider distribution reduces NLL at extremes. We therefore fix the per-window median of fully denoised samples and scale their deviations until their IQR equals that of the halted output. This counterfactual preserves center and width but lacks the tail-shape and multiscale spectral changes induced by stable convolution. It separates distributional shrinkage from undifferentiated scale inflation.

\method{} beats the equal-width counterfactual in 61 of 68 interior-optimum cells, with the clearest separation in the heavy-tailed region. Matching width therefore does not reproduce the tail shape or multiscale suppression of stable convolution.

\begin{figure}[htbp]
\centering
\includegraphics[width=\columnwidth]{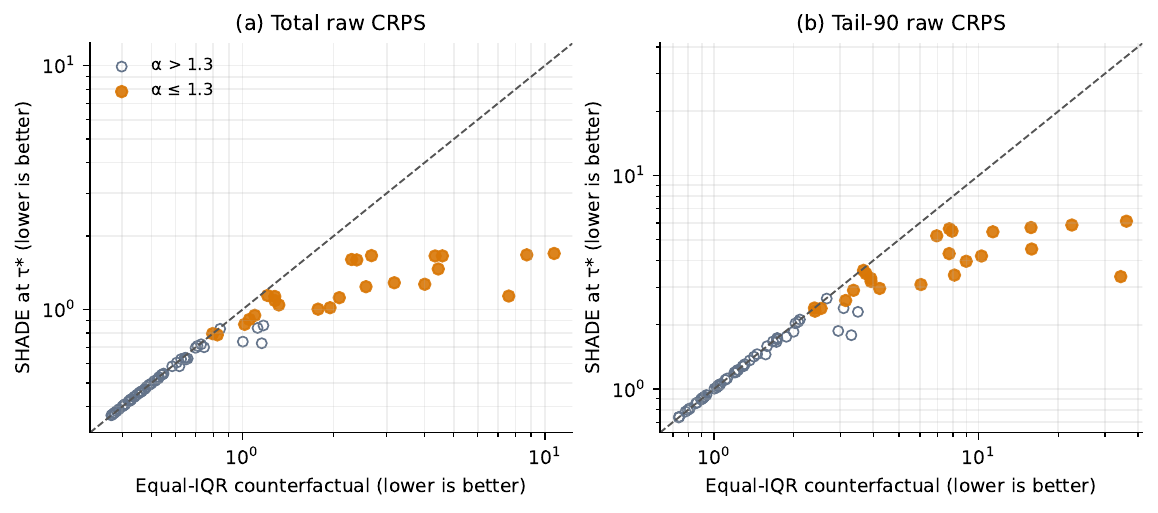}
\caption{Paired comparison between stable halted shrinkage and equal-IQR variance inflation.}
\label{fig:equal-iqr}
\end{figure}

\subsubsection{Noise-Family Misspecification}

Real disturbances need not be stable: they may have finite variance, conditional heteroscedasticity, or volatility clustering. We therefore generate observation noise from Student-$t(\nu=3)$ and GARCH(1,1) processes while fixing the diffusion model at $\alpha_{\mathrm{model}}=1.5$. This deliberate mismatch tests whether halting exploits a general endpoint-estimation failure rather than the closed form of the stable generating family.

For Student-$t$ noise at $\mathrm{SNR}=0.1$, raw CRPS changes only from 0.705 to 0.704, while mean NLL falls from 20.61 to 1.25. For GARCH noise at $\mathrm{SNR}=0.5$, CRPS changes from 0.574 to 0.573, NLL falls from 15.90 to 1.21, and 90\% coverage moves from 0.83 to 0.91. The optimal interior depth varies across families and SNRs, but the common pattern is recovery of tail density and coverage with little change in average CDF error. Thus, the mechanism does not require exact stable observation noise; equally, these results should not be read as improvement under every score. Complete six-cell results and depth curves are reported in the appendix.

\subsubsection{Components and Mechanisms}

Together, the tail-index scan isolates the applicable regime, LIM isolates the effect of stable noise alone, equal-IQR removes scale inflation, misspecification removes exact family matching, and the two halting routes test the need for explicit scale labels. The component ordering shows that stable diffusion alone is insufficient: retaining a data-dependent residual kernel produces the main correction, and context amortization adapts it across windows. Full scans results and alternative components experiment results appear in the appendix.

\section{Conclusion}

\method{} treats diffusion depth as a learnable regularization path: a nonzero readout retains a stable low-pass kernel that can reduce finite-sample endpoint error. FiLM conditions the denoiser on tail thickness, and scale-calibrated or context-amortized halting selects the shrinkage strength. Synthetic, misspecified-noise, equal-width, and real-data evidence consistently localizes the benefit to heavy-tail density calibration rather than universal improvement in average error.

This interpretation turns diffusion time from a computational coordinate into statistical resolution and makes halting part of the estimator, not a sampling shortcut. The local power law is specific to the stable semigroup and does not automatically transfer to other generators. Moreover, our theory calibrates marginal conditional CDFs, not multistep or multivariate joint tails. Extending structured halting to energy or variogram scores and asymmetric decision losses is therefore the main next step.

\bibliography{references}

@inproceedings{rasul2021timegrad,
  title={Autoregressive Denoising Diffusion Models for Multivariate Probabilistic Time Series Forecasting},
  author={Rasul, Kashif and Seward, Calvin and Schuster, Ingmar and Vollgraf, Roland},
  booktitle={Proceedings of the 38th International Conference on Machine Learning},
  pages={8857--8868},
  year={2021}
}

@inproceedings{tashiro2021csdi,
  title={CSDI: Conditional Score-Based Diffusion Models for Probabilistic Time Series Imputation},
  author={Tashiro, Yusuke and Song, Jiaming and Song, Yang and Ermon, Stefano},
  booktitle={Advances in Neural Information Processing Systems},
  volume={34},
  pages={24804--24816},
  year={2021}
}

@inproceedings{yuan2024diffusionts,
  title={Diffusion-TS: Interpretable Diffusion for General Time Series Generation},
  author={Yuan, Xinyu and Qiao, Yan},
  booktitle={International Conference on Learning Representations},
  year={2024}
}

@inproceedings{kim2021noise2score,
  title={Noise2Score: Tweedie's Approach to Self-Supervised Image Denoising without Clean Images},
  author={Kim, Kwanyoung and Ye, Jong Chul},
  booktitle={Advances in Neural Information Processing Systems},
  volume={34},
  pages={864--874},
  year={2021}
}

@inproceedings{yoon2023levy,
  title={Score-Based Generative Models with L{\'e}vy Processes},
  author={Yoon, Eunbi and Park, Keehun and Kim, Sungwoong and Lim, Sungbin},
  booktitle={Advances in Neural Information Processing Systems},
  volume={36},
  year={2023}
}

@article{gneiting2007proper,
  title={Strictly Proper Scoring Rules, Prediction, and Estimation},
  author={Gneiting, Tilmann and Raftery, Adrian E.},
  journal={Journal of the American Statistical Association},
  volume={102},
  number={477},
  pages={359--378},
  year={2007}
}

@book{samorodnitsky1994stable,
  title={Stable Non-Gaussian Random Processes},
  author={Samorodnitsky, Gennady and Taqqu, Murad S.},
  publisher={Chapman and Hall},
  year={1994}
}

@article{kalita2024aneeg,
  title={AnEEG: Leveraging Deep Learning for Effective Artifact Removal in EEG Data},
  author={Kalita, Bhabesh and Deb, Nabamita and Das, Daisy},
  journal={Scientific Reports},
  volume={14},
  pages={24234},
  year={2024},
  doi={10.1038/s41598-024-75091-z}
}

@inproceedings{wang2024financial,
  title={A Financial Time Series Denoiser Based on Diffusion Models},
  author={Wang, Zhuohan and Ventre, Carmine},
  booktitle={Proceedings of the 5th ACM International Conference on AI in Finance},
  pages={1--9},
  year={2024},
  doi={10.1145/3677052.3698649}
}

@article{liu2024manufacturing,
  title={A Causal Based Method for Denoising Non-Homologous Noises in Time Series Manufacturing Monitoring Data},
  author={Liu, Changqing and Li, Yingguang and Hua, Jiaqi and Zhao, Zhiwei and Gao, James},
  journal={Journal of Manufacturing Systems},
  volume={76},
  pages={92--102},
  year={2024},
  doi={10.1016/j.jmsy.2024.07.008}
}

@inproceedings{zhou2024decl,
  title={Denoising-Aware Contrastive Learning for Noisy Time Series},
  author={Zhou, Shuang and Zha, Daochen and Shen, Xiao and Huang, Xiao and Zhang, Rui and Chung, Korris},
  booktitle={Proceedings of the Thirty-Third International Joint Conference on Artificial Intelligence},
  pages={5644--5652},
  year={2024},
  doi={10.24963/ijcai.2024/624}
}

@inproceedings{guo2025dydiff,
  title={Dynamical Diffusion: Learning Temporal Dynamics with Diffusion Models},
  author={Guo, Xingzhuo and Zhang, Yu and Chen, Baixu and Xu, Haoran and Wang, Jianmin and Long, Mingsheng},
  booktitle={International Conference on Learning Representations},
  year={2025}
}

@inproceedings{ye2025nsdiff,
  title={Non-Stationary Diffusion for Probabilistic Time Series Forecasting},
  author={Ye, Weiwei and Xu, Zhuopeng and Gui, Ning},
  booktitle={Proceedings of the 42nd International Conference on Machine Learning},
  volume={267},
  pages={72112--72130},
  year={2025}
}

@inproceedings{james1961estimation,
  title={Estimation with Quadratic Loss},
  author={James, William and Stein, Charles},
  booktitle={Proceedings of the Fourth Berkeley Symposium on Mathematical Statistics and Probability},
  volume={1},
  pages={361--379},
  year={1961}
}

@article{yao2007early,
  title={On Early Stopping in Gradient Descent Learning},
  author={Yao, Yuan and Rosasco, Lorenzo and Caponnetto, Andrea},
  journal={Constructive Approximation},
  volume={26},
  pages={289--315},
  year={2007}
}

@article{moran2018eegsnr,
  title={Improving the Quality of a Collective Signal in a Consumer {EEG} Headset},
  author={Mor{\'a}n, Alejandro and Soriano, Miguel C.},
  journal={PLOS ONE},
  volume={13},
  number={5},
  pages={e0197597},
  year={2018},
  doi={10.1371/journal.pone.0197597}
}

@article{herreraparedes2018communities,
  title={Design of Synthetic Bacterial Communities for Predictable Plant Phenotypes},
  author={Herrera Paredes, Sur and Gao, Tianxiang and Law, Theresa F. and Finkel, Omri M. and Mucyn, Tatiana and Teixeira, Paulo Jos{\'e} Pereira Lima and Salas Gonz{\'a}lez, Isa{\'i} and Feltcher, Meghan E. and Powers, Matthew J. and Shank, Elizabeth A. and Jones, Corbin D. and Jojic, Vladimir and Dangl, Jeffery L. and Castrillo, Gabriel},
  journal={PLOS Biology},
  volume={16},
  number={2},
  pages={e2003962},
  year={2018},
  doi={10.1371/journal.pbio.2003962}
}

@article{lim2024semgdataset,
  title={Assessment of Self-Report, Palpation, and Surface Electromyography Dataset During Isometric Muscle Contraction},
  author={Lim, Jihoon and Lu, Lei and Goonewardena, Kusal and Liu, Jefferson Zhe and Tan, Ying},
  journal={Scientific Data},
  volume={11},
  number={1},
  pages={208},
  year={2024},
  doi={10.1038/s41597-024-03030-8}
}

@article{cont2001stylized,
  title={Empirical Properties of Asset Returns: Stylized Facts and Statistical Issues},
  author={Cont, Rama},
  journal={Quantitative Finance},
  volume={1},
  number={2},
  pages={223--236},
  year={2001},
  doi={10.1080/713665670}
}

@article{aitsahalia2005microstructure,
  title={How Often to Sample a Continuous-Time Process in the Presence of Market Microstructure Noise},
  author={A{\"i}t-Sahalia, Yacine and Mykland, Per A. and Zhang, Lan},
  journal={The Review of Financial Studies},
  volume={18},
  number={2},
  pages={351--416},
  year={2005},
  doi={10.1093/rfs/hhi016}
}

@article{langarica2023blind,
  title={Contrastive Blind Denoising Autoencoder for Real-Time Denoising of Industrial {IoT} Sensor Data},
  author={Langarica, Sa{\'u}l and N{\'u}{\~n}ez, Felipe},
  journal={Engineering Applications of Artificial Intelligence},
  volume={120},
  pages={105838},
  year={2023},
  doi={10.1016/j.engappai.2023.105838}
}

@article{zhang2024ssl,
  title={Self-Supervised Learning for Time Series Analysis: Taxonomy, Progress, and Prospects},
  author={Zhang, Kexin and Wen, Qingsong and Zhang, Chaoli and Cai, Rongyao and Jin, Ming and Liu, Yong and Zhang, James Y. and Liang, Yuxuan and Pang, Guansong and Song, Dongjin and Pan, Shirui},
  journal={IEEE Transactions on Pattern Analysis and Machine Intelligence},
  volume={46},
  number={10},
  pages={6775--6794},
  year={2024},
  doi={10.1109/TPAMI.2024.3387317}
}

@inproceedings{zhang2024llmts,
  title={Large Language Models for Time Series: A Survey},
  author={Zhang, Xiyuan and Chowdhury, Ranak Roy and Gupta, Rajesh K. and Shang, Jingbo},
  booktitle={Proceedings of the Thirty-Third International Joint Conference on Artificial Intelligence},
  pages={8335--8343},
  year={2024},
  doi={10.24963/ijcai.2024/921}
}

@article{raskutti2014early,
  title={Early Stopping and Non-Parametric Regression: An Optimal Data-Dependent Stopping Rule},
  author={Raskutti, Garvesh and Wainwright, Martin J. and Yu, Bin},
  journal={Journal of Machine Learning Research},
  volume={15},
  number={11},
  pages={335--366},
  year={2014}
}

@article{gneiting2011point,
  title={Making and Evaluating Point Forecasts},
  author={Gneiting, Tilmann},
  journal={Journal of the American Statistical Association},
  volume={106},
  number={494},
  pages={746--762},
  year={2011},
  doi={10.1198/jasa.2011.r10138}
}

@inproceedings{zeng2023linear,
  title={Are Transformers Effective for Time Series Forecasting?},
  author={Zeng, Ailing and Chen, Muxi and Zhang, Lei and Xu, Qiang},
  booktitle={Proceedings of the AAAI Conference on Artificial Intelligence},
  volume={37},
  number={9},
  pages={11121--11128},
  year={2023},
  doi={10.1609/aaai.v37i9.26317}
}

@article{alexandrov2020gluonts,
  title={GluonTS: Probabilistic and Neural Time Series Modeling in Python},
  author={Alexandrov, Alexander and Benidis, Konstantinos and Bohlke-Schneider, Michael and Flunkert, Valentin and Gasthaus, Jan and Januschowski, Tim and Maddix, Danielle C. and Rangapuram, Syama and Salinas, David and Schulz, Jasper and Stella, Lorenzo and T{\"u}rkmen, Ali Caner and Wang, Yuyang},
  journal={Journal of Machine Learning Research},
  volume={21},
  number={116},
  pages={1--6},
  year={2020}
}

@inproceedings{nie2023patchtst,
  title={A Time Series Is Worth 64 Words: Long-Term Forecasting with Transformers},
  author={Nie, Yuqi and Nguyen, Nam H. and Sinthong, Phanwadee and Kalagnanam, Jayant},
  booktitle={International Conference on Learning Representations},
  year={2023}
}

@article{nachmani2021nongaussian,
  title={Non-Gaussian Denoising Diffusion Models},
  author={Nachmani, Eliya and San Roman, Robin and Wolf, Lior},
  journal={arXiv preprint arXiv:2106.07582},
  year={2021}
}

@article{deasy2021heavytail,
  title={Heavy-Tailed Denoising Score Matching},
  author={Deasy, Jacob and Simidjievski, Nikola and Li{\`o}, Pietro},
  journal={arXiv preprint arXiv:2112.09788},
  year={2021}
}

@inproceedings{pandey2025heavytail,
  title={Heavy-Tailed Diffusion Models},
  author={Pandey, Kushagra and Pathak, Jaideep and Xu, Yilun and Mandt, Stephan and Pritchard, Michael and Vahdat, Arash and Mardani, Morteza},
  booktitle={International Conference on Learning Representations},
  year={2025},
  url={https://openreview.net/forum?id=tozlOEN4qp}
}

@inproceedings{simsekli2017fractional,
  title={Fractional {L}angevin {M}onte {C}arlo: Exploring {L}{\'e}vy Driven Stochastic Differential Equations for {M}arkov Chain {M}onte {C}arlo},
  author={{\c{S}}im{\c{s}}ekli, Umut},
  booktitle={Proceedings of the 34th International Conference on Machine Learning},
  series={Proceedings of Machine Learning Research},
  volume={70},
  pages={3200--3209},
  year={2017}
}

@inproceedings{shariatian2025dlpm,
  title={Heavy-Tailed Diffusion with Denoising {L}{\'e}vy Probabilistic Models},
  author={Shariatian, Dario and {\c{S}}im{\c{s}}ekli, Umut and Durmus, Alain Oliviero},
  booktitle={International Conference on Learning Representations},
  year={2025},
  url={https://openreview.net/forum?id=SYmUS6qRub}
}

@inproceedings{sun2022discrete,
  title={Score-Based Continuous-Time Discrete Diffusion Models},
  author={Sun, Haoran and Yu, Lijun and Dai, Bo and Schuurmans, Dale and Dai, Hanjun},
  booktitle={International Conference on Learning Representations},
  year={2023},
  note={arXiv preprint first released in 2022},
  url={https://openreview.net/forum?id=BYWWwSY2G5s}
}

@inproceedings{chen2023jump,
  title={Learning to Jump: Thinning and Thickening Latent Counts for Generative Modeling},
  author={Chen, Tianqi and Zhou, Mingyuan},
  booktitle={Proceedings of the 40th International Conference on Machine Learning},
  series={Proceedings of Machine Learning Research},
  volume={202},
  pages={5367--5382},
  year={2023}
}

@article{baule2025jumpdiffusion,
  title={Generative Modelling with Jump-Diffusions},
  author={Baule, Adrian},
  journal={arXiv preprint arXiv:2503.06558},
  year={2025}
}

\end{document}